# Learning Personalized Discretionary Lane-Change Initiation for Fully Autonomous Driving Based on Reinforcement Learning

Zhuoxi Liu
*Department of Mechanical Engineering*
*The University of Tokyo*
Tokyo, Japan
t-ryuu@iis.u-tokyo.ac.jp

Zheng Wang
*Institute of Industrial Science*
*The University of Tokyo*
Tokyo, Japan
z-wang@iis.u-tokyo.ac.jp

Bo Yang
*Institute of Industrial Science*
*The University of Tokyo*
Tokyo, Japan
b-yang@iis.u-tokyo.ac.jp

Kimihiko Nakano
*Institute of Industrial Science*
*The University of Tokyo*
Tokyo, Japan
knakano@iis.u-tokyo.ac.jp

*Abstract*—In this article, the authors present a novel method to learn the personalized tactic of discretionary lane-change initiation for fully autonomous vehicles through human-computer interactions. Instead of learning from human-driving demonstrations, a reinforcement learning technique is employed to learn how to initiate lane changes from traffic context, the action of a self-driving vehicle, and in-vehicle user's feedback. The proposed offline algorithm rewards the action-selection strategy when the user gives positive feedback and penalizes it when negative feedback. Also, a multi-dimensional driving scenario is considered to represent a more realistic lane-change trade-off. The results show that the lane-change initiation model obtained by this method can reproduce the personal lane-change tactic, and the performance of the customized models (average accuracy 86.1%) is much better than that of the non-customized models (average accuracy 75.7%). This method allows continuous improvement of customization for users during fully autonomous driving even without human-driving experience, which will significantly enhance the user acceptance of high-level autonomy of self-driving vehicles.

*Keywords—autonomous driving, lane-change initiation, personalized model, reinforcement learning, human-machine interface*

## I. Introduction

Autonomous vehicles are expected to be prevalent within the next decade, which have the potential to reduce the number of road fatalities due to human error and improve productivity in cars. Much more than that, there is an increasing demand for better user experience while autonomous driving. A lot of studies suggest that driving styles differ from individuals [1]. An excellent human-centered self-driving system should provide a safe, reliable, and user-adaptive experience.

In an autonomous driving system shown in Fig. 1, the decision-making and planning module [2] serves as the 'brain' that receives and analyzes information from the perception module and generates decisions for the executors. Sequential planning or hierarchical control is widely used as the tool to construct the decision-making module for self-driving vehicles. The three-level hierarchy is a well-accepted structure to characterize the hierarchical decision-making of human drivers, which was first proposed in 1985 [3] and has been investigated for driver modeling by many prior studies [4][5]. The three levels are strategic level, tactical level, and operational level, which are summarized in TABLE I. The strategic level is concerned with long-term or large-scale decisions such as route planning and mode selection. Tactical control refers to the planning of certain maneuvers to achieve short-term objectives, for example, to decide whether to follow or overtake the leading car through the interactions with other road users and infrastructures. The operational level covers direct vehicle-controlling operations such as steering and gearing. The upper level not only encompasses the lower level in timescale but also influences the lower level. Some studies modeled these layers separately, in this case, the output of the upper level is the input of the lower level.

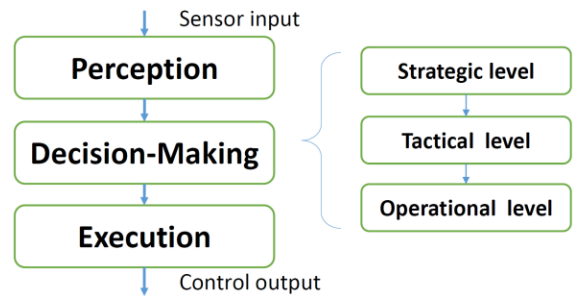

Fig. 1. Control process of an autonomous driving system.

TABLE I. Hierarchical Control of Driving

| Control level | Hierarchy | Timescale | Examples |
|---|---|---|---|
| Strategic | 1 | Long-term (minutes /hours) | Route planning; mode choice; |
| Tactical | 2 | Short-term (seconds) | Turns, lane changes, stops; |
| operational | 3 | Critical (Hundreds of milliseconds) | Steering, press pedal, and gearing; |

This research is supported by a Grant-in-Aid for Early-Career Scientists (No. 19K20318) from the Japan Society for the Promotion of Science.

*A. Decision-Making for Lane-Change*

Lane change is one of the most frequent driving behaviors on the road. 10% of freeway crashes occur during lane changes [6]. Most previous works that modeled lane-change behavior heavily focused on the operational level, which aimed at generating a comfortable and efficient lane-change trajectory [4][7]. For example, how to generate a lane-change trajectory and execute it when the decision of initiating a lane change has already been made. While tactical behavioral planning did not gain so much attention as the operational driving [8]. Tactical lane change decision is more about the determination of lane change, i.e. where and when to initiate a lane change, which is prior in time and order to the operational lane change control. Answers to how to generate tactical decisions are often arbitrary and subjective. Most of the researchers treated lane change as an alternative for collision avoidance, where lane changes are forced to happen due to emergency cases, such as possible collision with a leading car. However, most of the lane changes are naturally initiated due to the driver's desire during the interactions with other road users. The former type of lane change is defined as *mandatory lane change* and the latter one is called *discretionary lane change* [9].

The decision-making model for lane change can be usually categorized into physical model and statistical model. A physical model is process-driven where the mathematical laws and processes are pre-defined. In [10] and [11], the authors used physical models to determine parametric trajectories that optimize the time-saving and energy-saving conditions respectively. However, the physical model cannot deal with unexpected or unconsidered situations and sometimes requires heavy manual tuning. A statistical model is a data-driven model, where the model is estimated based on data without explicit knowledge of the physical behavior of the system. In [12][13], the trajectories were learned from real driving data using machine learning techniques. The statistical method becomes an emerging trend because it can not only leverage big data to intelligently learn some rules, but also tackle with diverse real-life driving situations and various personalized demands. Concerning user-adaptive modeling, the learning-based approaches are more efficient to be realized than the process-driven ones.

*B. Related studies*

Two studies are found to model discretionary lane-change initiation based on drivers' demonstrations [14][15]. In [14], the authors described such a discretionary lane when the vehicle is approaching a preceding vehicle whose speed is slower than itself. They proposed an autonomous lane change initiation and control structure where both tactical and operational level were incorporated. For the tactical decision, they used a Support Vector Machine, a supervised learning algorithm, to determine whether to continue lane-keep or initiate a lane change, where the classifier was trained with the data of actual human drivers' demonstrations. In [15], lane-change behavior is deconstructed as the adjustment of longitudinal position, gap acceptance to initiate the lane change and operations of the lane change. The authors used linear regression, also a supervised learning algorithm, to achieve personalized models from three subjects.

In studies [14][15], personalized lane-change tactics were learned from human drivers' on-their-own lane-change demonstrations, which relied on the method called *Supervised Learning (SL)* in machine learning. Intuitively speaking, this method means that human drivers teach the 'machine' how initiate lane changes by their own correct demonstration, in other words, the 'machine' directly learns the tactics from the correct answers. However, cloning human-driving behaviors for a fully autonomous driving vehicle may be too arbitrary and sometimes not easy to achieve. First, in fully autonomous vehicles, there is technically no driver. So there no chance for the intelligent agent to learn a 'driver model' from a person's previous driving experience. What is more, human-driving vehicle behaviors have intrinsic characteristics which are different from a self-driving car's. For example, the cognitive process of humans causes a latency (sometimes called reaction time) between perception and execution, drivers' behaviors, such gap distance maintenance in the car-following task, is based on their cognitive properties. While as the counterpart of the cognitive process of the human, the decision-making process of self-driving cars relies on computer-based architecture which means there is no human cognitive latency. If we involve such a human cognitive latency happening while human is driving in an autonomous driving system, we may sacrifice the efficiency of the system. In general, improving the customization of a fully autonomous driving system is still in demand and therefore personalized modeling without direct human-driving demonstration is necessary for fully autonomous vehicles.

In this study, we employ *Reinforcement Learning (RL)* to provide an alternative solution to personalizing discretionary lane-change initiation for autonomous driving. In this method, a learning agent learns a certain tactic from a trial-and-error game, where the in-vehicle user gives his feedback as the rewards for the learning agent's behavior and the learning agent learns from its mistake. To our knowledge, such a method has never been used to obtain a personalized lane-change initiation model so far.

*C. Purpose*

Finding a good opportunity to initiate a lane change is always a trade-off in many circumstances. Lane-change initiation depends on not only the contextual cues (e.g. kinematics of other road users, infrastructure) but also the decisions of prior importance (i.e. higher-level decision). In this research, we consider a discretionary lane change situation in an expressway where there is a slower preceding car that gets in the way of the user car while the user car is required to driving autonomously at a higher speed. This situation can be well-interpreted by the three-level hierarchy structure, where the strategic decision of the user car is being in a time-saving mode (i.e. driving at a high speed to minimize the total time consumption to the destination), then the tactical reasoning is hereby confined. We propose a logic for decision-making on whether lane-keep or lane change in this situation, which is illustrated in Fig. 2. The lane change will be initiated unless some certain condition is satisfied. In addition to that, the vehicle will keep the lane and continuously measure the distance to the preceding car. The threshold is set for safety. If the gap is equal to or lower than the threshold, the vehicle will keep a safe gap to the leading car to avoid a collision. The threshold is determined by the neutral drivers' expected headway with considerations of personalization in the car-following driver model [16]. In the case of driving at a speed of 80km/h, the threshold is calculated as 40 meters.

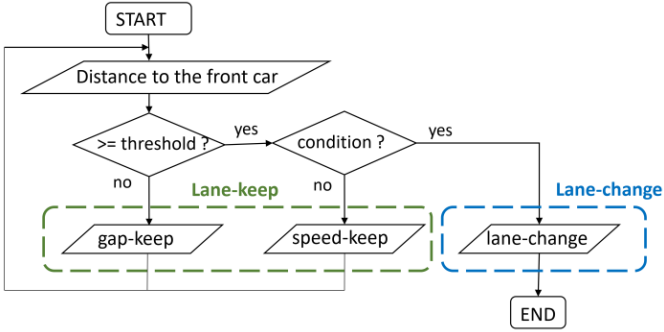

Fig. 2. Decision-making logic for discretionary lane change when there is a slower preceding vehicle.

We assume that a user has his own acceptable lane change condition. Therefore, our task in this research is to learn personalized 'condition' from different individuals in the situation of fully autonomous driving and to validate our assumptions. The customized model is supposed to be able to represent a person's intentions to some extent. The more the users' actual intentions can be reproduced by the lane-change initiation model, the higher user acceptance will be.

Based on the discussion on the related works, instead of using the supervised learning technique where the human-driving data are directly learned, the reinforcement learning technique is utilized to reproduce a person's accepted lane-change initiation conditions. We take advantage of a keyboard-based human-machine interface and leverage the feedback directly given by in-vehicle users in a fully autonomous situation. Also, we consider a multi-dimensional context in a discretionary lane-change situation. Our solution is more suitable and realistic for human-centered improvements in fully autonomous driving.

The purposes of this research are: (1) to propose a reinforcement-learning-based method to derive the personalized tactical decision-making model for lane-change initiation with the consideration of multi-dimensional driving context; and (2) to validate this method through simulated driving experiments. This article is organized as follows. Section II presents the formal definitions of the reinforcement learning problem and the proposed algorithm. Section III describes how the data used to train and test the model are collected in a simulated autonomous driving experiment. The results are given and discussed in Section IV. Section V summarizes this research finally.

## II. METHOD

Reinforcement Learning (RL) is one of Machine Learning (ML) paradigms concerning an agent's learning of a certain behavior through trial-and-error interactions with a dynamic environment [17]. The agent in RL does not directly learn from explicit examples of behaviors, instead, it explores the environment and updates its knowledge by judging the rewards or punishments given by the environment in a repetitive game. In a standard RL problem, for each trial, the agent chooses an action under a given state and receives a reward for its behavior. Moreover, its behavior will affect what state it will transit into in the next trial.

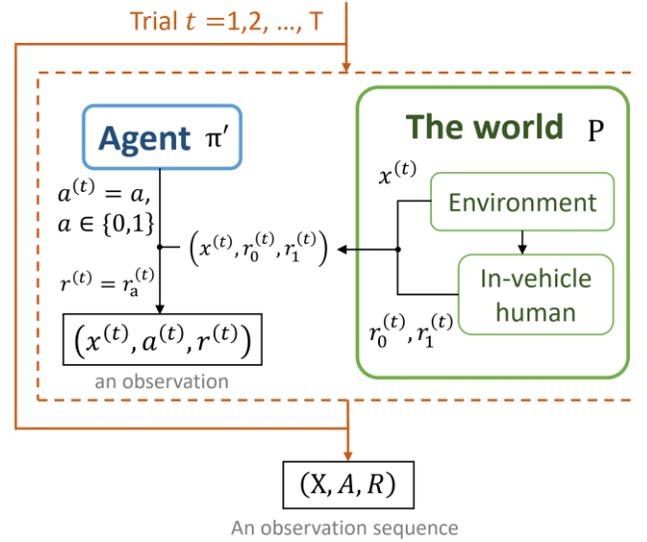

Fig. 3. Offline contextual bandit problem.

K-Armed *Contextual Bandit (CB)* problem is an extension of the RL problem, where the 'context' refers to the states and k arms of the bandit to pull means k available actions to choose. Basically, in each trial of a CB problem, an arm is pulled in a certain context and consequently, a reward is rendered. Different from the RL problem, the arms pulled in the previous trial will not lead to a certain state in the next trial, which means there is no transition between states. Please note that the terms and notations of CB are different from that of RL. In this part, we use expressions in the field of the bandit problem to define and formulate our target problem and algorithm.

### A. Problem Formulation

In this work, the offline learning of discretionary lane-change initiation from human feedback in a series of traffic situations can be framed as an offline Contextual two-armed Bandit, where the 'context' refers to the differing traffic situations and 'two arms' represent the decisions of action: lane change and lane keep.

An offline CB problem, as shown in Fig. 3, is a game where an agent tries to obtain a sequence of observations with a policy $\pi'$ through interaction with the world. Based on the definition of a standard CB problem [18], the offline CB problem in this work is formally defined as follows: $a \in \{0,1\}$ is the arm to pull in the discretionary lane-change situation, where $a = 0$ and $a = 1$ mean choosing lane change and lane keep. There is a distribution $P$ over $(X, R_0, R_1)$, where $X$ is the context of the driving scenario, and $R_a \in \{-1,1\}$ is the reward for pulling arm $a$. On the trial $t$ of the repetitive CB game as seen in Fig. 3, a sample point $(x^{(t)}, r_0^{(t)}, r_1^{(t)})$ is drawn from $P$, where $x^{(t)}$ is the context which is a partial representation of the environment, i.e. driving scenario, $r_0^{(t)}, r_1^{(t)}$ are the rewards given by in-vehicle human for pulling arm $a = 0$ and $a = 1$ in the context $x^{(t)}$. The agent pulls an arm $a^{(t)}$ based on its policy $\pi' \sim Bernoulil(p = 0.5)$, and the corresponding reward $r^{(t)} = r_{a^{(t)}}^{(t)}$ is then revealed. Therefore, from the trial $t$, an observation $(x^{(t)}, a^{(t)}, r^{(t)})$ is obtained. From an experiment consisting of $T$ trials, a sample $\{(x^{(1)}, a^{(1)}, r^{(1)}), (x^{(2)}, a^{(2)}, r^{(2)}), \cdots, (x^{(T)}, a^{(T)}, r^{(T)})\}$ will be taken. We denote such a sample set as $(X, A, R)$ in this research, where $X = \{x^{(1)}, ..., x^{(T)}\}$ , $A = \{a^{(1)}, ..., a^{(T)}\}$ , $R = \{r^{(1)}, ..., r^{(T)}\}$, $T$ is the size the this set.

*B. Algorithm*

In this research, we employ an off-policy CB algorithm as the solver to the learning of human-like lane-change initiation from the observation sequence in the CB problem.

Our objective is to learn the policy π to pull the best arm that receives the maximum expected reward. The policy $\pi = \pi(a|x)$ is the mapping from action $a$ and context $x$. Note that the policy $\pi'$ to pull an arm in the CB problem is not the policy π that we aim to evaluate and optimize, that is how 'off-policy' comes. The policy π is the target policy that we want to learn from the data sample, which reflects the decision-making strategy of the human observer who provides feedback in the CB problem. In such a setting the learning agent only has access to a prior sampled offline dataset of experience. Some researchers argue that in many real-world RL applications, direct online interactions with the environment is limited, while previously collected dataset of logged experience is more accessible [19]. Therefore, we firstly experiment with an off-policy CB to leverage the existing data.

*1) Optimization Goal*

The goal of the CB algorithm is to find an optimal policy for the agent to obtain maximum rewards. This algorithm target modeling and optimizing the policy directly. The policy is usually modeled with a parameterized function respect to $\theta$, $\pi_\theta(a|s)$. The objective function sums up all the rewards and its value is based on the policy. So, an algorithm can be applied to optimize $\theta$ for the maximum value of the objective function. The objective function is defined as follows.

$$J(\theta) = \rho + \sum_t \pi_\theta(a^{(t)}|x^{(t)}) \cdot r^{(t)} \quad (1)$$

Where $a^{(t)}, x^{(t)}$, and $r^{(t)}$ are the action, context, and reward at trial $t$, $\pi_\theta(a|x)$ is the policy that decides which action to take given context $x$, $r^{(t)} \in \{-1,1\}$ is the reward for pulling arm $a^{(t)}$ in context $x^{(t)}$. $\rho$ is a regularization term added to prevent overfitting. The optimization goal is to maximize the objective function, i.e. $\max_\theta J(\theta)$.

We use an artificial neural network (ANN) to model the policy $\pi_\theta(a|x)$. The input of ANN is the context $x$, the dimension of $x$ depends on how many factors we choose to represent the context, which will be discussed in detail in Section III. The output layer has two units with sigmoid function as the activation. The sigmoid function exists from 0 to 1. So, the two output units are interpreted as the probabilities of receiving a positive reward for $a = 0$ and $a = 1$ respectively.

Intuitively speaking, when the user gives a positive feedback $r^{(t)} = 1$ the optimization goal is equivalent to $\max_\theta \pi_\theta(a^{(t)}|x^{(t)})$, which means rewarding the policy for pulling $a^{(t)}$ in $x^{(t)}$; while if it is a negative feedback $r^{(t)} = -1$, the goal equals to $\min_\theta \pi_\theta(a^{(t)}|x^{(t)})$, which means penalizing the policy for pulling $a^{(t)}$ in $x^{(t)}$.

*2) Pseudocode of the CB algorithm*

We use the *Batch Gradient Descent* as the optimizer to update our objective function. The pseudocode of the CB algorithm is given in Fig. 4.

The training set $\{X, A, R\}$, validation set $\{S^{val}, A^{val}, R^{val}\}$, and test set $\{S^{test}, A^{test}\}$ are the data set collected in autonomous driving experiments, which will be introduced in detail in Section III. The batch size is the number of elements in the batch $\{X', A', R'\}$, which is set as 32. And the training is stopped automatically by examining the training process, that is if the validation performance becomes stable, the training is done. We use accuracy as the metric to evaluate the model. When computing accuracy during training, we compare the estimated $A$ with the 'true $A$'. The 'true $A$' is the action that the in-vehicle human will take in the context $X$ according to his policy π, The 'true $A$' is directly given as $A^{test}$ in test set $\{X^{test}, A^{test}\}$, but it is not directly given in the training set $\{X, A, R\}$, where $A$ is the action taken by the agent according to $\pi'$), so we assume the 'true $A$' by considering $A$ and $R$ both: when $r = 1$ and $a = 0$ *or* 1, 'true $a$' = 1 or 0; when $r = 0$ and $a = 0$ *or* 1, 'true $a$' = 1 or 0.

---

**Load data**: Training set $\{X, A, R\}$ , and test set $\{X^{test}, A^{test}\}$;
**Split** a validation set $\{X^{val}, A^{val}, R^{val}\}$ from training set;
**Initialize**: trainable parameters $\theta$ in the ANN;
**Iterate**: for each epoch $e \in \{1, \cdots, E\}$:
    - Draw a random batch $\{X', A', R'\}$ from $\{X, A, R\}$;
    - Update the policy: $\theta \leftarrow \max_\theta J(\theta)$ with $\{X', A', R'\}$ by *Gradient Descent* (learning rate = 0.1);
    - Estimate $A^{val}$ given $X^{val}$ with the policy and compute validation accuracy;
    - Estimate $A$ given $X$ with the policy compute training accuracy;
    **If** Standard deviation validation of validation accuracies from 1000 most recent steps < 0.01 (only if $t < 1000$)
        - stop the iteration;
    **Else**
        - continue the iteration;
**Evaluation**: estimate $A^{test}$ given $X^{test}$ with the policy and

Fig. 4. Offline Contextual bandit algorithm (pseudocode).

## III. DATA COLLECTION AND EXPERIMENT

The data used for training and evaluation are collected in a simulated autonomous driving experiment. In this section, we will introduce the experimental setting, driving scenario, and details about the collected data. The experiments were done in a driving simulator (DS), as seen in Fig. 5. Four subjects participated in the experiment.

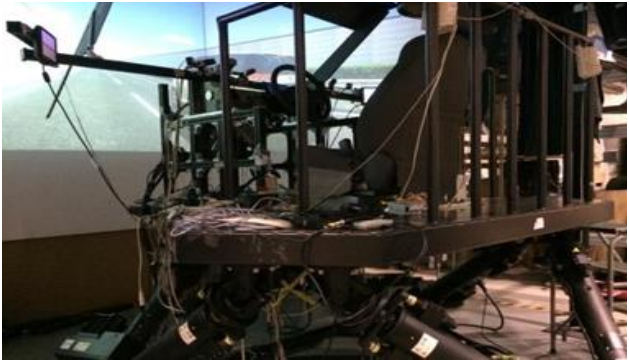

Fig. 5. DS experiment setting.

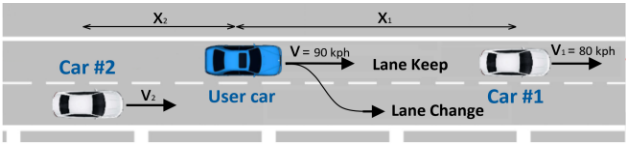

Fig. 6. Driving scenario.

TABLE II. INDEPENDENT VARIABLES

| Independent variables | Description | Level |
|---|---|---|
| $x_1$ | Gap from car #1 to user car | {40, 50, 60, 70, 80} (m) |
| $x_2$ | Gap from user car to car #2 | {10, 20, 30, 40, 50, 60} (m) |
| $x_3$ | Velocity of car #2 | {80, 90,100} (kph) |
| $a$ | User car's action | {lane change, lane keep} |

The user car in DS is programmed to driving autonomously on a two-lane expressway. The driving scenario is illustrated in Fig. 6, where another two cars are traveling in front of the user in the user lane and behind the user in the adjacent lane respectively. The velocities of user car and car #1 are set to 90kph and 80kph, and the priority of the user car is maintaining its speed, which induces a non-emergency lane-change situation. Besides, four independent variables were considered in such a scenario which are summarized in TABLE Ⅱ. The number of levels of an independent variable is the number of experimental conditions. The first three variables in TABLE Ⅱ are used to describe the context and the last one denotes the user car's action. Although 'three/two-second rule', a rule that requests drivers to keep at least three or two seconds headway time, is suggested in many countries to help drivers to keep safe headways and avoid collisions, many drivers still keep their habits of distancing leading vehicles and usually the time headway is shorter than the suggested. Therefore, the variable $x_1$ is sampled based on a survey of actual headway time distribution on a national highway of Japan, where 67.1% of the observed vehicles fall into 0.5s~2.5s time headway, and the interval of 1.5s~2.0s has the majority which is the 19.5% [20]. We sample the most frequent gap distance from the preceding car according to the distribution of headway time on the highway. It is noteworthy that either the actual headway time or suggested headway time is under the human-driving situation, which might be different from that under a self-driving situation.

This is because, for example, self-driving cars can avoid latency due to the reaction time of human driving. Therefore, we assume that users have higher acceptance of lower time headway when the car is autonomous. In this experiment, by imposing the concept that the car is driving autonomously and has its mechanism to avoid collisions, the subjects' mindsets are in a fully autonomous diving situation.

An experiment is divided into two sessions for collecting the training data and test data respectively. In the first session for collecting training data, a subject went through all the possible combinations of chosen independent variables and gave their feedback to the user car's action through a keyboard. In the second session for collecting test data, a subject designates his own choices of actions given a series of contexts.

The first session consists of episodic scenarios with varying context and action. In each episode, a scenario with a certain combination of independent variables discussed above is exhibited by the DS to a subject. The subject is firstly informed of the user car's decision by an icon displayed on the screen, right at the same moment, the subject starts to visually observe the car #1 and car #2 from the front window and right side-view mirror respectively. Three seconds after that the user car then starts to practice its decision. The subject's feedback is either negative or positive, which depends on whether he agrees with the user's car's decision in such a context. One episode lasts for 16 seconds totally, which allows enough time for the subject to provide their response. Overall, each subject experienced around 180 different episodes, which allows them to go through as many contexts as possible. In this session, a data set of context, action, and reward is obtained for each subject. We draw a proportion of data collected in this part as the validation dataset. The rest is used as the training dataset. The interface that collects subjects' feedback is the keyboard where there are only two keys available which represent 'yes' and 'no' (i.e. positive and negative feedback) respectively.

The second session is, in general, the same as the first session except for that the actions of the user car are not determined. Instead, the subjects directly designated the action via the winker according to their own intentions.

## IV. RESULT AND DISCUSSION

### A. Feedback Consistency

High-quality data is the foundation of successful machine learning. To confirm the reliability of a subject's performance in the experiment, the consistency of their feedback was examined. In the experiment, the action of the user car is binary in the scenario (i.e. lane change or lane keep), which means that in the same context, a subject must agree with only one of the two actions. If a subject gives positive feedback upon lane change in a context, he ought to hold negative feedback with lane keep in the same context, and vice versa. If a subject gives the same feedback in two trials where the contexts are the same, but the actions are opposite, the subject is being self-contradictory, and such feedback will be considered as being inconsistent. We define the consistency ratio as the ratio of trials with consistent feedback to the total trials. The consistency ratios of four subjects then are calculated, which are 89%, 86.7%, 78.7%, and 44%.

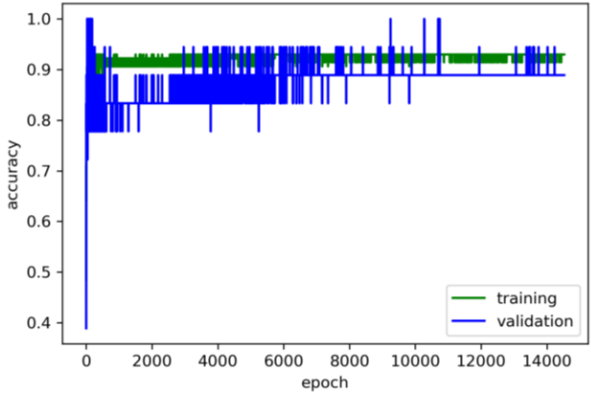

Fig. 7. Learning Curve.

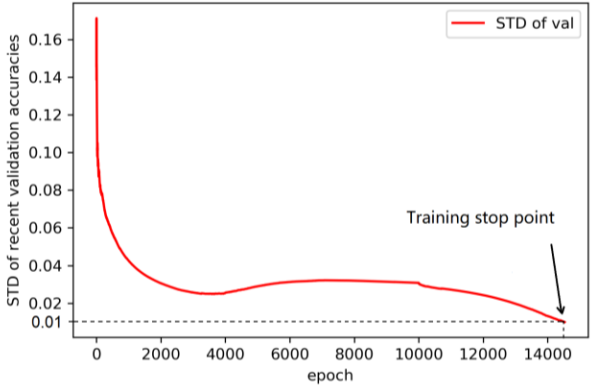

Fig. 8. Standard deviation of validation accuracies.

We assume that there should be some feedback inconsistencies existing in some contexts where a participant is not very decisive in some contexts, but such cases should not be the majority. Learning data with poor quality may not lead to a satisfactory result. Therefore, we reject the data of the subject with a 44% consistency ratio and only experiment with our algorithm on the data of the other three subjects.

### B. Model Selection and Training

A learning curve, as shown in Fig. 7, illustrates how the learning agent performs (measured on the horizontal axis) over trials of training (measured on the vertical axis). The performance is evaluated by training score (i.e. the estimation accuracy on training data) and validation score (i.e. the estimation accuracy on validation data). The training process can be monitored through the learning curve. When the validation accuracy becomes stable and smooth, the training is then completed. We use the standard deviation (STD) of the most 1000 recent validation accuracies over the epochs as a metric to measure the stability of validation performance, which is shown in Fig. 8. As training proceeds, the STD becomes smaller. The training is stopped while the validation accuracy is above 80% and the STD value is lower than 0.01.

The hyperparameters are the parameters that are not trainable. The values of them must be determined before the final training process. The hyperparameters in our model are the architecture of ANN (i.e. number of hidden layers, units of each hidden layers, activation function), the regularization term in the objective function, and the learning rate of the optimizer. These hyperparameters are tuned carefully. The optimal combination of hyperparameters is the one that gives the maximum test accuracy and needs fewer epochs to complete the learning. Finally, one hidden layer with four units without activation is set, the regularization term is set to be 1, and the learning rate is 0.1.

### C. Training Results and Evaluation

Three models are trained with three subjects' training sets respectively. We then test these models with the subjects' test set one by one. The test results are shown in TABLE III. We termed a model customized model when it is tested with the same person's data. The values on the diagonal of TABLE III. are the test accuracies of customized models for subject #1, 2, and 3 respectively, and the rest ones are that of non-customized models. We can observe that for any subject, the performance of the customized model is always better than that of non-customized models. We compare them by a scatter plot as seen in Fig. 9. The average accuracies of the customized model and the non-customized model are 86.1% and 75.7% respectively. It is obvious to see that the test accuracies of customized models are higher and have less dispersion than the non-customized models, which indicates that the customized model has better and more stable performance than the non-customized model for an individual.

TABLE III.  MODEL TEST ACCURACIES

| Model<br>Test set | Subject #1 | Subject #2 | Subject #3 |
|---|---|---|---|
| Subject #1 | 0.8541 | 0.5625 | 0.8542 |
| Subject #2 | 0.75 | 0.8333 | 0.875 |
| Subject #3 | 0.8541 | 0.6458 | 0.8958 |

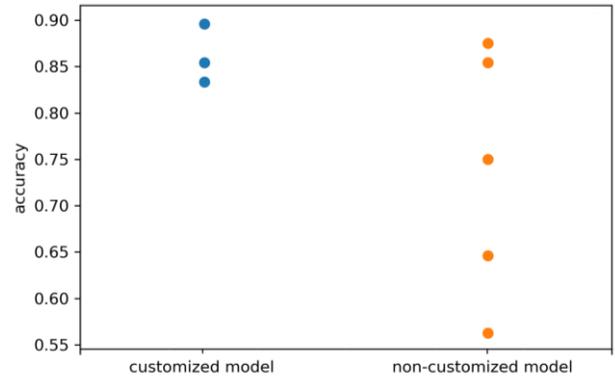

Fig. 9. Performance of customized model vs non-customized model

### D. Comparison with Related Studies

We compare 3 studies (including ours) with the same research purpose in TABLE IV. All of these works aim at learning personalized lane-change initiation tactics from users of autonomous vehicles using machine learning techniques. It is noteworthy that the performance of these methods cannot be precisely compared due to different driving scenarios and test metrics. But all of the studies achieve satisfactory results by their methods. This study differs greatly from the others in the method and algorithm. The authors of [14] and [15] relied on SL while this work creatively employ RL to address the problem. Our solution enables human-demonstration-free personalized modeling which can be applied not only for driver assistance driving (SAE level: L1) and semi-autonomous driving (L2, L3) but also for high and fully autonomous driving (L4, L5).

TABLE IV. COMPARISON WITH RELATED STUDIES

|  | **This Study** | **[14]** | **[15]** |
|---|---|---|---|
| Method | RL | SL | SL |
| Algorithm | Contextual Bandits | Support Vector Machine | Linear Regression |
| Experimental setting | Simulated | On-road | On-road |
| Driving scenario | Three-vehicle scenario | Three-vehicle situation | Four-vehicle situation |
| Subject | Three | Two | Three |
| Model performance | 86.1% (average accuracy) | 6% 7% for two subjects (average normalized error) | Being able to capture personalizations |
| Human's driving | No need | Need | Need |
| Applied SAE level | L1 ~ L5 | L1, L2, L3 | L1, L2, L3 |

## V. CONCLUSION

In this article, we proposed a novel reinforcement learning method to learn multi-dimensional personal tactic of discretionary lane-change initiation in the context of fully autonomous driving. Unlike most of the machine learning solutions that learn personalities from real human-driving experience, this method can be carried out in day-to-day fully autonomous driving where there is no human driver but the users are still allowed to continuously improve and personalize their own cars by a simple interface. We tested our off-line Contextual Bandits algorithm on three subjects and found that the performance of the customized model is significantly better than the non-customized model, which proves that this method can distinguish personal traits of the decision-making of lane-change initiation in autonomous driving. In the future, we will extend this method in the real-time setting and test it in more complex and diverse driving scenarios